# THE MASS DISTRIBUTION IN CLUSTERS OF GALAXIES FROM WEAK AND STRONG LENSING

JORDI MIRALDA-ESCUDÉ
*Institute for Advanced Study*
*Olden Lane, Princeton NJ 08540, USA*
*Email: jordi@sns.ias.edu*

## 1. Abstract

This paper is intended as an introduction to the theory of weak lensing. A review of the inversion formula introduced by Kaiser and Squires is presented. We then prove the formula of the aperture densitometry method in a simple way that allows a clear understanding of where the various terms come from. This is particularly useful to quantitatively measure masses in any region of a lens. We then summarize what has been learned from observations of strong lensing about the dark matter distribution; weak lensing should provide similar information on larger scales in clusters of galaxies.

## 2. The Weak Lensing Inversion

Gravitational lensing can be described as a mapping from the image plane (with coordinates $x_I$, $y_I$) to the source plane (with coordinates $x_S$, $y_S$), which is given by the lens equation:

$$(x_S, y_S) = (x_I, y_I) - \nabla\phi(x_I, y_I) , \qquad (1)$$

where $\nabla\phi$ is the deflection angle (in the single screen approximation), the projected gravitational potential $\phi$ is related to the surface density $\Sigma$ by $\nabla^2\phi = 2(\Sigma/\Sigma_{crit})$, and $\Sigma_{crit}$ is the critical surface density (see Schneider, Ehlers, & Falco 1992). The distortion of a small image is given by the magnification matrix, which is the Jacobian of this mapping:

$$A^{-1} \equiv \frac{\partial(x_S, y_S)}{\partial(x_I, y_I)} \equiv \begin{pmatrix} 1 - \kappa - \lambda & -\mu \\ -\mu & 1 - \kappa + \lambda \end{pmatrix} . \qquad (2)$$

Using the lens equation, the convergence can be expressed as $\kappa = 1/2\,\nabla^2\phi = \Sigma/\Sigma_{crit}$, while the two components of the shear are

$$\lambda = \frac{1}{2}\left(\frac{\partial^2\phi}{\partial^2 x_I} - \frac{\partial^2\phi}{\partial^2 y_I}\right)\ , \qquad \mu = \frac{\partial^2\phi}{\partial x_I\,\partial y_I}\ . \qquad (3)$$

The magnification matrix can be diagonalized by rotating the coordinates by some angle $\beta$, to the form

$$A^{-1} = \begin{pmatrix} 1-\kappa-\gamma & 0 \\ 0 & 1-\kappa+\gamma \end{pmatrix}\ , \qquad (4)$$

where $\lambda = \gamma\cos(2\beta)$, and $\mu = \gamma\sin(2\beta)$. The axes in this rotated frame are called the "principal axes of the shear", because the images are "stretched" along these axes. The "stretching factor" $q$, or the axis ratio of the image of a circular source, is the ratio of the two eigenvalues $q = (1-g)/(1+g)$, where $g = \gamma/(1-\kappa)$. Thus, the change in the ellipticities of background sources depends only on the quantity $g$. In the limit of weak lensing, we assume $\gamma \ll 1$ and $\kappa \ll 1$, so $q \simeq 2\gamma$ and the observed quantity from the galaxy ellipticities is the shear. The problem of weak lensing is then reduced to obtaining the surface density given the shear.

To solve this problem, we first see how the shear is expressed in terms of the surface density. The projected potential is a linear superposition of the potentials caused by every element of mass in the lens:

$$\phi(\mathbf{r}_I) = \int d\mathbf{r}'_I\,\frac{\Sigma(\mathbf{r}'_I)}{\pi\Sigma_{crit}}\log\|\mathbf{r}_I - \mathbf{r}'_I\|\ , \qquad (5)$$

where $\mathbf{r}_I$ is the vector $(x_I, y_I)$ (this results from the fact that $\nabla^2\log\|\mathbf{r}_I - \mathbf{r}'_I\| = 2\pi\,\delta^2(\mathbf{r}_I - \mathbf{r}'_I)$ ). Using equation (3), the shear is given by

$$[\lambda(\mathbf{r}_I), \mu(\mathbf{r}_I)] = \int d\mathbf{r}'_I\,\frac{\kappa(\mathbf{r}'_I)}{\pi}\,\frac{-[\cos 2\eta, \sin 2\eta]}{\|\mathbf{r}_I - \mathbf{r}'_I\|^2}\ . \qquad (6)$$

Here, the quantities written inside square parentheses are "polars" which, like the shear, rotate by twice the angle by which the coordinates are rotated. We have defined $\eta$ as the angle between the direction from $\mathbf{r}_I$ to $\mathbf{r}'_I$ and the x-axis. With the minus sign, the orientation of the polar inside the integral rotates to the tangential direction which the shear at $(x_I, y_I)$ produced by a mass element at $(x'_I, y'_I)$ should have.

The inversion of the operator in equation (6) is most easily found by considering the Fourier transforms of the convergence, the shear and the potential, which we denote as $\kappa_k$, $[\lambda_k, \mu_k]$, and $\phi_k$, and will be functions of the two Fourier coordinates $k_x, k_y$. The second derivatives giving the convergence and the shear in terms of the potential are converted into multiplications with $k_x$ and $k_y$ in Fourier space, so equation (6) is simply given by:

$$[\lambda_k, \mu_k] = \left[ \frac{k_x^2 - k_y^2}{k_x^2 + k_y^2}, \frac{2k_x k_y}{k_x^2 + k_y^2} \right] \kappa_k . \tag{7}$$

It is most easily seen that the scalar product of the operator in equation (7) with itself is equal to unity. In other words, this operator is its own inverse. If true in Fourier space, this must also be true in real space, where the operator has the form in equation (6). Thus, the surface density is expressed in terms of the shear as

$$\kappa(\mathbf{r}_I) = \int d\mathbf{r}_I', \frac{[\lambda(\mathbf{r}_I'), \mu(\mathbf{r}_I')]}{\pi} \cdot \frac{-[\cos 2\eta, \sin 2\eta]}{\|\mathbf{r}_I - \mathbf{r}_I'\|^2} . \tag{8}$$

This equation was found by Kaiser & Squires (1993). As they explained, it is obviously not directly applicable since the surface density can only be recovered with finite resolution. Several methods to obtain convolved maps, solving the problem of the finite size of the fields, and extending the analysis beyond the weak lensing limit have been extensively discussed (Kaiser 1995; Schneider & Seitz 1995; Seitz & Schneider 1995; Bartelmann 1995). Here, we shall give an alternative proof of the inversion equation which gives us directly the surface density convolved with a particularly useful window function, in the aperture densitometry method.

The main operation in equation (8) is to calculate an average of the tangential component of the shear (which is obtained from the scalar product of the shear with the polar $[\cos 2\eta, \sin 2\eta]$) around every point, as proposed originally by Tyson, Valdes, & Wenk (1990). We can then simply do this average over a circle for a point mass lens, which can only depend on the distance from the point mass to the center of the circle. Since any arbitrary lens can be decomposed into small mass elements that can be treated as point masses, this will give us an inversion formula for the surface density smoothed with a certain filter. Using the symbols defined in Figure 1, the tangential component of the shear at a point along the circle is $\lambda_t(\phi) = (b/D)^2 \cos 2\beta$, where $b$ is the critical radius of the point mass, and using simple trigonometry, the average along the circle $\lambda_t^C$ is

$$\lambda_t^C = \frac{b^2}{R^2} \int_0^{2\pi} \frac{d\phi}{2\pi} \frac{(1 - a\cos\phi)^2 - (a\sin\phi)^2}{(1 - a\cos\phi)^2 + (a\sin\phi)^2} . \tag{9}$$

where $a = d/R$. The remarkable result is that this integral is equal to 1 for $a < 1$, and equal to zero for $a > 1$. Thus, *any mass outside the circle has no effect on $\lambda_t^C$, while any mass which is inside has an effect which is independent of its position within the circle*. In fact, $b^2/R^2$ is the average convergence within the circle, $\bar{\kappa}$, due to the point mass (since, by definition, the average surface density within the critical radius is equal to the critical surface density). Thus, for any arbitrary lens, the contribution from the mass inside the circle is $\lambda_t^C = \bar{\kappa}$. We still need to include the contribution from the mass *on* the circle. We first notice that, from symmetry considerations, the contribution from any element of mass cannot depend on its azimuthal angle, and therefore the contribution from the mass on the circle can only depend on the surface density averaged over the circle, $\kappa^C$. Since we also know that the shear is zero when the surface density is uniform, the contribution from the mass on the circle must be $-\kappa^C$, so the result for any arbitrary lens is

$$\lambda_t^C = \bar{\kappa} - \kappa^C \ . \tag{10}$$

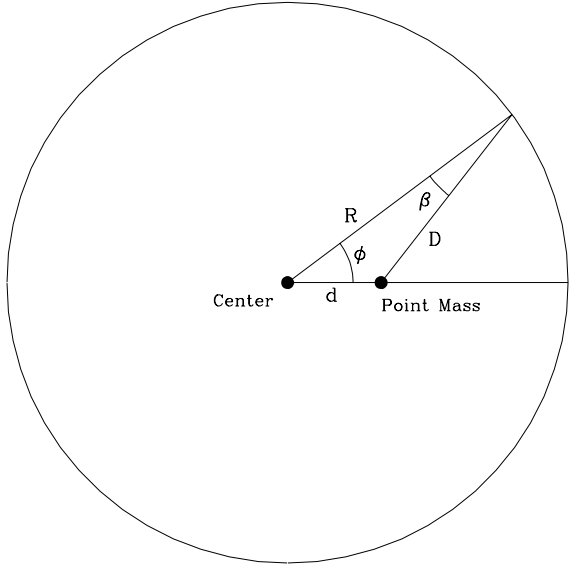

*Figure 1.*

This equation was known for a circularly symmetric lens (where $\lambda_t^C = \gamma(r)$ is simply the total shear), but its general validity was not realized until recently. The integration of this equation over an interval in $\log R$, where $R$ is the radius around a fixed center, yields (using $\bar{\kappa}(R) = 2/R^2 \int dR' \, R' \, \kappa^C(R')$ and exchanging the order of the integrals)

$$2 \int_{R_1}^{R_2} d\log R \ \lambda_t^C(R) = \bar{\kappa}(R_1) - \bar{\kappa}(R_2) \ , \tag{11}$$

which gives the surface density convolved with a compensated top-hat (see Fahlman et al. 1994, Kaiser 1995). This formula is useful not only to measure mass differences on the largest scale of an observed frame, but also to obtain maps of the small-scale variations in surface density (which could indicate the presence of clumpiness in the dark matter) by applying the formula around all possible centers.

## 3. Observations of Strong Lensing and Future Prospects

After weak lensing was first detected by Tyson et al. (1990), it is only recently that a large number of observational results are being reported (Bonnet, Mellier, & Fort 1994; Fahlman et al. 1994; Smail, Ellis, & Fitchett 1994; Smail et al. 1995; Tyson & Fisher 1995). Weak lensing by individual galaxies, first searched by Tyson et al. (1984), is now also being found (Brainerd, Blandford, & Smail 1995), and the average shear in blank fields (the weakest lensing in the sky) is also being searched for (Mould et al. 1994; Villumsen 1995). The weak shear is only sensitive to differences in surface density, as is clearly seen from the inversion formulae extended only over finite fields (10), (11). Recent work by Broadhurst, Taylor, & Peacock (1995) has highlighted the use of the magnification, measured from the fluctuations in the number counts of galaxies in different colors, which has the advantadge of probing directly the surface density although it may be more severely subject to systematic errors due to the clustering of the lensed sources.

Meanwhile, much progress has been done on strong lensing, the observations of the highly distorted "arcs" and multiple images. After the initial realization that this ruled out a flat core larger than $\sim 50\,h^{-1}$ Kpc for the dark matter, it has been found that lensing always occurs around bright central cluster galaxies or around highly compact galaxy clumps, and the mass-to-light ratios are typically $M/L_B \sim 300\,h$ within the radius of the arcs (Fort & Mellier 1994). Models reproducing the multiply imaged sources require the presence of substructure in the form of dark matter clumps around the brightest galaxies; furthermore, these clumps are found to have similar ellipticities as the outer stellar isophotes of the bright galaxies (see Kneib et al. 1993, 1995; Mellier et al. 1993). This is not surprising, since the outer stellar isophotes can be observed to radii similar to those of the observed arcs, and therefore they must be in orbits in the dark matter clumps yielding the lensing deflection angles; it also implies that the stellar velocity dispersions of the bright galaxies must rapidly rise within the radius of the arcs (Miralda-Escudé 1995). These observations suggest that clusters of galaxies are typically not in a static equilibrium, but they are constantly undergoing mergers on a wide range of scales, down to the scales of the observed arcs and the stellar halos of the central galaxies. This picture seems consistent with hierarchical theories of formation of clusters and should offer new clues on the formation of central cluster galaxies (Merritt 1985 and references therein). The determination of masses can also provide new information on the hot intracluster gas (Loeb & Mao 1994, Miralda-Escudé & Babul 1995, Waxman & Miralda-Escudé 1995, Allen, Fabian, & Kneib 1995, Squires et al. 1995). The present rapid growth of weak lensing observations will probably lead to new discoveries on similar aspects of clusters of galaxies at larger radii: the mass-to-light ratios and the density profiles, substructure in the dark matter and its relation to the galaxy distribution, and the physical state of the intracluster gas.

I thank Nick Kaiser for many illuminating discussions on this subject. I gratefully acknowledge support from the W. M. Keck Foundation.